# An investigation into the minimum number of tissue groups required for 7T in-silico parallel transmit electromagnetic safety simulations in the human head

Matthijs H.S. de Buck[1], Peter Jezzard[1], Hongbae Jeong[2,3], and Aaron T. Hess[4,5]

[1]Wellcome Centre for Integrative Neuroimaging, FMRIB Division, Nuffield Department of Clinical Neurosciences, University of Oxford, Oxford, United Kingdom, [2]Athinoula A. Martinos Center for Biomedical Imaging, Massachusetts General Hospital, Charlestown, MA 02129, USA, [3]Department of Radiology, Harvard Medical School, Boston, MA 02115, USA, [4]Oxford Centre for Clinical Magnetic Resonance Research, Department of Cardiovascular Medicine, University of Oxford, Oxford, United Kingdom, [5]BHF Centre of Research Excellence, University of Oxford, Oxford, United Kingdom

MANUSCRIPT SUBMITTED TO MAGNETIC RESONANCE IN MEDICINE

Word count: 3066/2800 (text body, excluding acknowledgements)

**Correspondence address**:

Aaron T. Hess, PhD

Oxford Centre for Clinical Magnetic Resonance Research

Department of Cardiovascular Medicine

University of Oxford

John Radcliffe Hospital

Oxford, OX3 9DU

United Kingdom

Email: aaron.hess@cardiov.ox.ac.uk

# Abstract

**Purpose:** Safety limits for the permitted Specific Absorption Rate (SAR) place restrictions on pulse sequence design, especially at ultra-high fields (≥7 tesla). Due to inter-subject variability, the SAR is usually conservatively estimated based on standard human models that include an applied safety margin to ensure safe operation. One approach to reducing the restrictions is to create more accurate subject-specific models from their segmented MR images. This study uses electromagnetic simulations to investigate the minimum number of tissue groups required to accurately determine SAR in the human head.

**Methods:** Tissue types from a fully characterized electromagnetic human model with 47 tissue types in the head and neck region were grouped into different tissue clusters based on the conductivities, permittivities, and mass densities of the tissues. Electromagnetic simulations of the head model inside a parallel transmit (pTx) head coil at 7T were used to determine the minimum number of required tissue clusters to accurately determine the subject-specific SAR. The identified tissue clusters were then evaluated using two additional well-characterized electromagnetic human models.

**Results:** A minimum of 4 clusters plus air was found to be required for accurate SAR estimation. These tissue clusters are centered around gray matter, fat, cortical bone, and cerebrospinal fluid. For all three simulated models the pTx maximum 10g-SAR was consistently determined to within an error of <12% relative to the full 47-tissue model.

**Conclusion:** A minimum of 4 clusters plus air are required to produce accurate personalized SAR simulations of the human head when using pTx at 7T.

**Keywords:** RF Safety, SAR simulation, ultra-high field, parallel transmit, subject-specific, clustered segmentation, k-means clustering

## Introduction

The energy deposition in human tissue permitted from radiofrequency pulses in MRI is limited by the energy deposition per unit mass, also known as the specific absorption rate or SAR. There are limits on both local SAR (averaged using 10g averaging volumes) and global SAR (1). Since SAR scales roughly with the square of the external magnetic field strength, accurate assessment of SAR is especially important for ultra-high field (≥7T) MRI-scanners. Due to individual differences in composition and morphometry of human anatomy, SAR varies across individuals for any given pulse. Additionally, local SAR cannot directly be measured in a clinical setting because of various technical and practical complexities.

Using computational human body models, it is possible to estimate SAR using electromagnetic simulations, such as with a finite-difference time-domain (FDTD) method (2,3). Those simulations determine, amongst other parameters, the electric field **E(r)** at each spatial location **r** in the model. This can be used to calculate the SAR at each location through:

$$SAR(\boldsymbol{r}) = \frac{1}{V} \int_{volume} \frac{\sigma(\boldsymbol{r})|\boldsymbol{E}(\boldsymbol{r})|^2}{2\rho(\boldsymbol{r})} d\boldsymbol{r} \qquad (1)$$

where V is the size of the averaging volume and $\sigma(\boldsymbol{r})$ and $\rho(\boldsymbol{r})$ are the conductivity and density of the tissue at location **r**, respectively.

These computational human body models are typically not available on an individual subject basis. Therefore, generic computational models have been used to determine the expected SAR, and then a further safety margin has been applied. Even when only considering a homogeneous group of adults, a safety margin of 1.5 is needed to correct for inter-subject variability with a chance of less than 1% of exceeding the calculated SAR in a pTx mode (4).

Parallel transmit MRI (5,6) has been shown to overcome $B_1$-inhomogeneities that are present at high field, and to reduce SAR (7). Although pTx enables spatial field manipulation, it can also create SAR 'hot-spots' by focusing the electric fields in an undesirable manner (6). This means that accurate subject-specific SAR is of particular interest when operating with pTx.

Previous work (8) studied the improvement of SAR simulations for ultra-high field pTx by non-linearly warping a standard electromagnetic model to match the anatomy of other individual subjects. That work demonstrated that morphometry alone is insufficient in determining accurate personalized SAR, but rather that personalized tissue composition must additionally be addressed.

Segmenting subject-specific MRI data has also been used to generate simplified anatomical models. Two examples which have been studied previously are models consisting of fat, lung, and water with uniform tissue densities for whole-body models at 3T (9), and fat, muscle, and skin images for prostate at 7T (10). In the head, previous work studied simulation results for several combinations of clusters based on anatomical proximity and similarity, but did not find cluster combinations that resulted in a stable estimation of SAR hotspots (11). More recent work has presented an automated segmentation approach using clusters consisting of tissues with similar tissue properties in multiple subjects, and evaluated the resulting SAR at 3T for only a single $B_1$-shim (12).

In this work we use a well-defined human electromagnetic model to evaluate the minimum number of tissue clusters needed to accurately estimate SAR for pTx in the head at 7T using a numerical clustering approach (13). The accuracy of SAR estimation using the identified tissue groups is then evaluated on two additional models with different ages and gender.

**Methods**

In this section we will first describe how the electromagnetic simulations were carried out and evaluated. We will then discuss how tissue clustering was achieved, and finally describe how an identified set of clusters was evaluated on different virtual human models.

Firstly, electromagnetic simulations were carried out for 7T (298 MHz) using models from the Virtual Population (IT'IS Foundation, Switzerland) (14) provided as part of the electromagnetic simulation package Sim4Life (Zurich MedTech, Switzerland). Simulations were performed initially for the original Duke model before additional simulations with altered tissue properties were done.. For all models, a portion of the shoulders was included to reduce the simulation time while avoiding problems at the boundary of the model (11,15). An optimized 8-channel pTx-coil (outer radius 146 mm, 5 tuning capacitors and 1 matching capacitor per channel, maximum coupling -11.3dB, no RF-shield) was also modelled. Multi-channel simulations were run by simulating each channel separately, with all other channels loaded with a 50 Ω load. The simulation setup with Duke in the pTx-coil is shown in Figures 1a-b. A non-uniform grid with an average voxel size of $(2.135 \text{ mm})^3$ and a maximum voxel size of $(3 \text{ mm})^3$ was used. All simulations were terminated after a convergence level of -30 dB was reached (as in (11)), which typically resulted in a simulated time of approximately 160 ns (or nearly 50 periods at 298 MHz). The simulations were run on a system using an Intel Xeon CPU E5-2680 (v4), running at 2.40 GHz with 14 cores and 28 logical processors. Running a single simulation using the voxelization and convergence described above typically took about six hours.

The 10g averaged SAR was calculated in Sim4Life using the IEEE/IEC 62704-1 standard (16). 10g-SAR maps were calculated for 64 different pTx configurations such that an 8×8-element Q-matrix (17) could be computed for every voxel in the model using the approach described by Beqiri *et al.* (18). Using the Q-matrices, each voxel's maximum 10g-SAR (for 1 W total input power) was determined through eigen-decomposition of the Q-matrices (19,20). Using this, the worst-case 10g-SAR can be defined as the highest maximum 10g-SAR for all voxels in the model for all possible $B_1$-shims. The maximum 10g-SAR for 500 random $B_1$ shims was also calculated, normalized to 1 W total input power per shim. Those 500 shims were generated by setting a random value for the power and phase for each channel, after which the powers were multiplied by a normalization factor such that the total input power was 1 W. Circular polarization (CP)-mode was included as an additional $B_1$-shim. For a given $B_1$-shim $\boldsymbol{w}_i$, the SAR corresponding to the Q-matrix of a voxel at location **r** is given by

$$SAR_i(\boldsymbol{r}) = \boldsymbol{w}_i^\dagger Q(\boldsymbol{r}) \boldsymbol{w}_i \qquad (2)$$

where the dagger (†) denotes the complex conjugate.

A list of 47 tissues from the head region was extracted from the Virtual Population v3.0 (ViP, IT'IS Foundation, Switzerland) (14). Amongst these 47 tissue types there were several with identical dielectric properties leading to only 41 unique tissue types. Tissues were grouped in a numerically optimized way based on their dielectric properties through k-means (21) clustering implemented using the *kMeans* function in the *scikit-learn* Python package (22). The k-means algorithm groups *n* vectors, using Euclidian distance, into *k* (with *k ≤ n*) clusters. For *k* = 1 to 6, clusters were identified using tissue conductivity and permittivity alone, setting the mass density to a fixed value of 1 g/cm$^3$ (as used in previous studies (9)), and then again with the additional inclusion of density (which is important when calculating SAR using Eqn 1). During clustering the *n* tissue types were weighted by their volumes in the original model. An example of k-means clustering with *k=5* is shown in Supporting Information Figure S1. For comparison, a 41-cluster model was also evaluated, representing the full model. An overview of all the *n* original tissue types can be found in Supporting Information Table S1.

Using the Duke model, simulations were set up for each tissue clustering method. Tissues within each cluster were assigned identical conductivity, permittivity, and density. The values were set to the corresponding value of the centroid of that cluster (i.e. the volume-weighted mean of the tissue properties in the cluster) in accordance with the results of the k-means clustering.

Using the identified minimum number of tissue clusters, the properties of 'real' tissues close to the cluster centroids were then identified and used to define tissue properties for that cluster. These tissue clusters and their properties were then evaluated for validity across different subjects in one simulation of Ella (female, 26y, 1.63 m), and one of Thelonious (male, 6y, 1.15 m), and using Duke in a different position (shifted 5 cm head-to-foot) and Duke with a smaller setting for the maximum voxel size (of 2.5 mm isotropic). For each model, the tissue clusters as identified for Duke from above were used and compared to simulation results using the original tissues from the respective member of the ViP.

For comparison, we assess the accuracy when using an automated MR-based segmentation-approach of the head-region into air, bone, fat, and soft tissues (as previously presented (12)). This closely resembles the '4 clusters plus air'-model presented here, but without the CSF-cluster.

**Results**

Figures 1c-e show the voxelized Duke-model with all 47 original tissue types.

Figure 2 shows the results of the simulations for the original version of Duke (full model), which function as the reference for all later simulations using simplified versions of Duke. The first row, Figures 2a-c, shows orthogonal projections of the overall maximums of the 10g-SAR per voxel. The highest values are towards the edge of the head close to the positions of the coils. In Figures 2d-f, the distribution for CP-mode is shown on a different color scale. Note that the maximum 10g-SAR for CP-mode is much lower than the worst-case 10g-SAR for all possible $B_1$-shims (0.140 W/kg versus 1.580 W/kg).

Figure 3 shows the errors in worst-case 10g-SAR for simulations with different numbers of k-means clusters, *k*. Results are shown for simulations with a simplified clustering, where a fixed mass density of 1 g/cm$^3$ is used for all tissues in the model, and for simulations using a different clustering, where the mass density is used as an additional clustering parameter. Even for *k=41* clusters, the simulations with unit density do not converge to the reference worst-case 10g-SAR, with a residual error of 19%. Therefore, the k-means clustering for determining the minimum required number of clusters was performed in a conductivity-permittivity-density hyperspace. Figure 3 shows that when using such a three-parameter clustering, both the worst-case 10g-SAR for all possible $B_1$-shims and the maximum 10g-SAR for the 500 random, normalized $B_1$-shims converge to close to the full model-values (errors of <1% for the worst-case 10g-SARs and 4.1±4.3% for the random shims) when using at least *k*=5 clusters based on the 3-parameter k-means clustering.

Based on this result, a clustering was made where the dielectric properties of five original tissues types were given to the different clusters. This approach was then used for subsequent validation simulations. One of these five clusters contained the internal air in the model, and the other clusters consisted of tissue types with centroids corresponding to gray matter, fat, cortical bone, and cerebrospinal fluid (CSF). Using this '4 clusters plus air' model, most of the simulated region is segmented to gray matter (8.2L out of 12.5L for Duke – 65.2%). 3.5L (27.7%) of the model is segmented to fat, 0.5L (4.2%) to cortical bone, 0.3L (2.1%) to CSF, and 0.1L (0.8%) to internal air. Supporting Information Table S1 shows which tissue types from the IT'IS model are assigned to each cluster.

The results for the '4 clusters plus air' model when used in a pTx mode are shown in Figure 4. Figures 4a-c show the voxelized Duke-model for this clustered segmentation. Figure 4d shows the ordered results for the 500 shims using Duke. Figure 4e shows the range of errors for all 500 shims in the five validation simulations (Duke, two additional ViP members, plus a z-shifted Duke and a version of Duke with altered resolution). In all cases there is an absolute error <12% for over 99% of the shims. The overestimate in worst-case 10g-SAR for Duke is 1.0%, for Ella it is 1.6%, and for Thelonious it is 0.3%. For all three models, the location of the worst-case SAR is in the same voxel for the '4 clusters plus air'-model as in their respective reference models. The simulations with a translated position of Duke and with a higher resolution simulation do not increase the error in 10g-SAR prediction.

Figure 5 shows the in-plane spatial distribution of errors in 10g-SAR for Duke when using the '4 clusters plus air' model instead of the full model. The results are shown for the three shims from the 500 random $B_1$-shims which correspond to the highest, median, and lowest 10g-SAR in the full Duke model. Figure 5 shows only the sagittal slices. The corresponding coronal and axial slices are shown in Supporting Information Figure S2.

Simulations using a previously presented imaging-based segmentation (without CSF) are shown in Figure 4e, which shows increased errors in the magnitude for the 500 random $B_1$-shims. The corresponding errors in magnitude and location of the overall maximum SAR compared to the simulations of the '4 clusters plus air'-model (including CSF), are shown in Supporting Information Table S3.

**Discussion**

When forming simplified tissue models, it was found that including mass density was important for the resultant model and for use as a clustering parameter. A tissue model simplified using a weighted 3D k-means clustering algorithm produces accurate 10g-SAR estimates for $k$=5 clusters in the head region at 7T. While using fewer clusters may make segmentation easier in a clinical setting, it may result in increased errors in both the simulated 10g-SAR for specific shims and the simulated worst-case 10g-SAR. Using $k$=5 clusters is both tractable to segment from *in-vivo* data and produces only small errors in 10g-SAR estimation.

Due to the weighting based on the volumes of the individual tissue types, the exact dielectric properties of the clusters in the $k$=5 k-means segmentation are model-dependent. Therefore, the results were studied for a segmentation using '4 clusters plus air' with the dielectric properties of 'real' tissues. The identified tissues (gray matter, fat, cortical bone, CSF, and internal air) were chosen because of their close resemblance to the dielectric properties of the tissues in the *k*=5 clusters k-means segmentation. This approach has the advantage that the properties of actual tissues are used in the IT'IS-database (23), which is not the case for the values returned by the k-means algorithm.

The resulting 10g-SAR-calculations, shown in Figures 4 and 5, exhibit a high degree of agreement with the reference data. This was found to be consistent for three human body models despite strongly different anatomies due to their size, weight, age, and gender. When expressed as a percentage of the shim-wise peak-values, the absolute errors are below 12% for over 99% of the shims for all models. Based on the results in Figures 4 and 5, the remaining 10g-SAR-errors generally seem to be over-estimations, which correspond to conservative SAR-estimations. The simulated errors are consistently much lower than the 50% uncertainty margin which is required for a probability of less than 1% of exceeding the actual value when using generic models. Also, the 50% margin is only sufficient when determining SAR for subjects of the same ethnicity as the generic model used for the simulation (e.g. if both model and subject are from the adult Caucasian population) (4). SAR simulation approaches using clustered segmentation, however, seem to offer consistent results for subjects of different genders and for both adults and children and may also be suitable for subjects with non-standard anatomies. Note that all results in this study are based on static $B_1$-shims – further analysis is required for more complex dynamic pTx-pulses.

The simulations were all carried out using the same 8-channel pTx coil. This coil was tuned *in silico* and constructed to minimize the influence of its design on the simulations, for example by using physically separated elements to minimize coupling between neighboring elements. The segmentation method has not yet been tested using coils with different designs.

All errors in the presented SAR simulations are based on models with perfect clustered segmentation. Generating practical clustered models for individual subjects would require experimental determination of the proposed subject-specific tissue clusters. Based on that segmentation, SAR simulations could then be conducted for individual subjects. The exact duration of SAR simulations depends on simulation parameters and computational capacity, but the computation time to date is not shorter than the duration of a typical MRI scan session. Therefore, in practice, the scan or scans which determine the clustered segmentation would likely have to be performed in a separate session, and therefore not necessarily at 7T, provided the resulting model can be correctly positioned in the pTx coil.

Recent work (12) used an automated computer-vision based approach that segmented the head regions of individual subjects into models consisting of air, bone, fat, and soft tissues for SAR simulation. Based on preliminary simulation results, they also concluded that the additional modelling of CSF is important for accurate SAR simulation, which agrees with our results. An augmented version of this approach, with CSF segmented as an additional cluster, would be interesting for the '4 clusters plus air' segmentation proposed here. However, the general lack of ground-truth information makes experimental validation challenging. Alternative automated segmentation approaches could make use of quantitative mapping of tissue properties such as the dielectric properties or the relaxation times (24, 25). In practice this would require high-accuracy quantitative maps of the whole head, for which the separation of bone and air is likely to prove difficult. Therefore, the identification of bone may have to be performed using a separate method, such as using ultra-short echo time scans (26) or a combination of T1- and T2-weighted images (27). Bone-air segmentation can be improved using additional post-processing steps, such as knowledge-based approaches (28) and artificial intelligence-based methods trained from CT (29), although such a network is not currently available for segmentation of the (upper part of the) skull at 3 tesla.

**Conclusion**

We found that a minimum number of 4 clusters plus air is required to generate personalized SAR models for the human head-region at 7T. A specific clustering approach is proposed whereby clusters are segmented to gray matter, fat, cortical bone, CSF, and air. This clustering resulted in errors in the simulated SAR-distributions and peak-10g-SAR values which are much smaller than the errors due to inter-subject variability when using generic models. The peak-10g-SAR could be determined with an error of less than 12% for models with different genders, age, and positioning in the scanner.

In order to be able to use this new approach in a clinical setting, an approach for the automated segmentation of the clusters in individual subjects, is still required. With that in place the newly proposed segmentation method could improve the estimation of subject-specific SAR, making it possible to operate 7T MR scanners closer to the true SAR-limits in a clinical setting.

**Acknowledgements**

The Wellcome Centre for Integrative Neuroimaging is supported by core funding from the Wellcome Trust (203139/Z/16/Z). We also thank the Dunhill Medical Trust and the NIHR Oxford Biomedical Research Centre for support (PJ). We acknowledge Zurich MedTech for help with Sim4Life. We also thank Mr Tony Zhou for helpful discussions. AH acknowledges support from the BHF Centre of Research Excellence, Oxford (RE/13/1/30181).

**Data availability statement**

In support of Magnetic Resonance in Medicine's reproducible research goal, the Python pipeline and the pTx coil model required for carrying out the simulations in Sim4Life as well as the relevant output-data have been made available online (doi: 10.5287/bodleian:nZGnByMav). The output data includes the Q-matrices and corresponding voxel-locations of all voxels in both the full model and the '4 clusters plus air' model for Duke, Ella, and Thelonious. The 500 random, normalized $B_1$-shims are also included.

**Figure Captions**

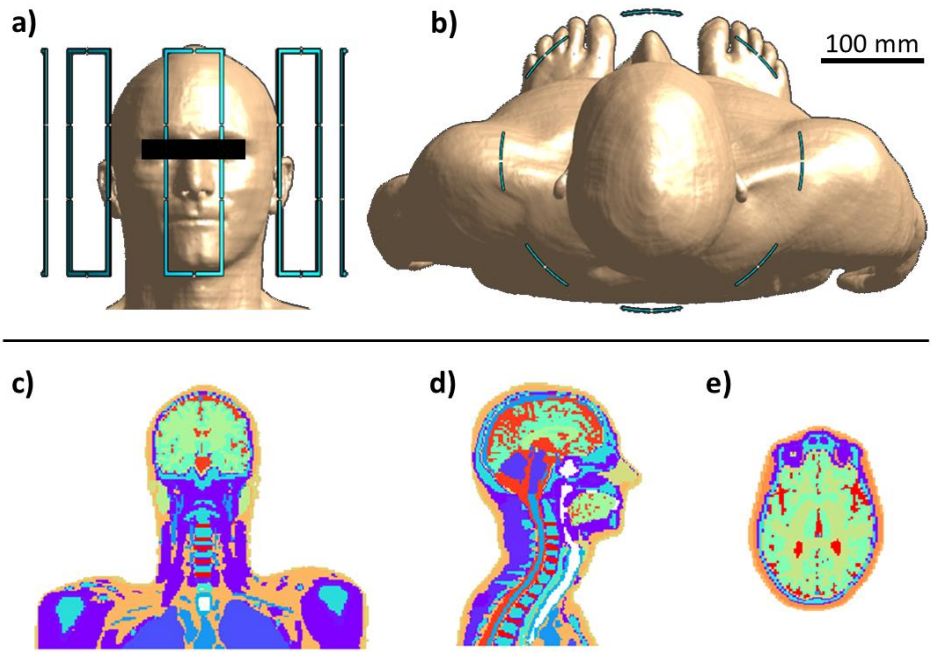

**Figure 1:** The simulation setup for Duke. **(a):** Front view of Duke positioned in the center of the 8-channel pTx-coil. **(b):** Top view of Duke in the coil. **(c-e):** Three orthogonal slices of the voxelized Duke model for the original segmentation with 47 tissue types (arbitrary colors used for different tissue types).

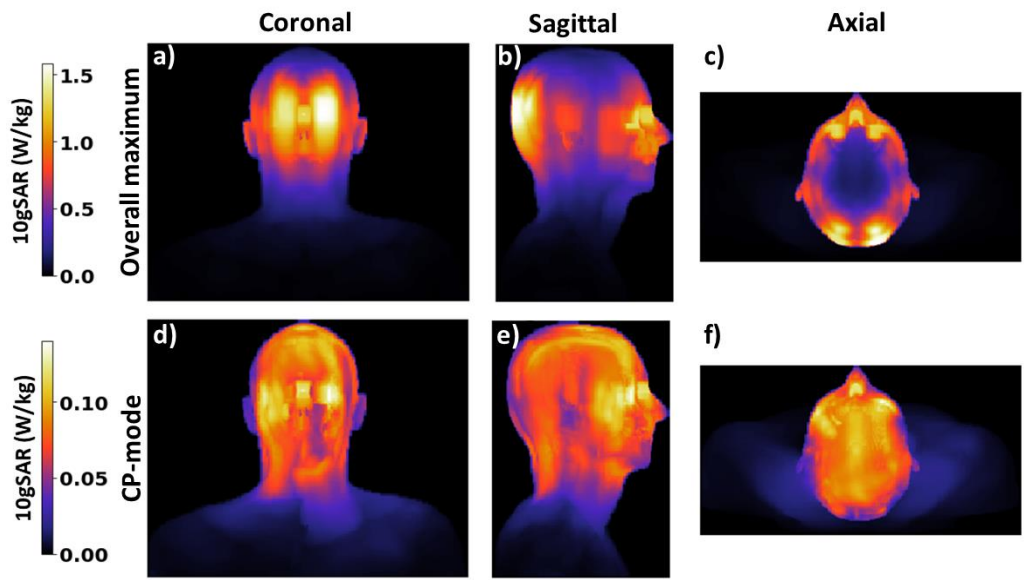

**Figure 2:** Ground-truth simulation results for the full Duke model. **(a-c):** Maximum intensity projections of the 10g-SAR-distribution in the full Duke model for the overall voxel-wise worst-case values of the 10g-SAR for all possible $B_1$-shims. **(d-f):** Maximum intensity projections of the 10g-SAR-

distribution in circular polarization (CP)-mode. Note the difference in color-bar scaling with figures (a-c).

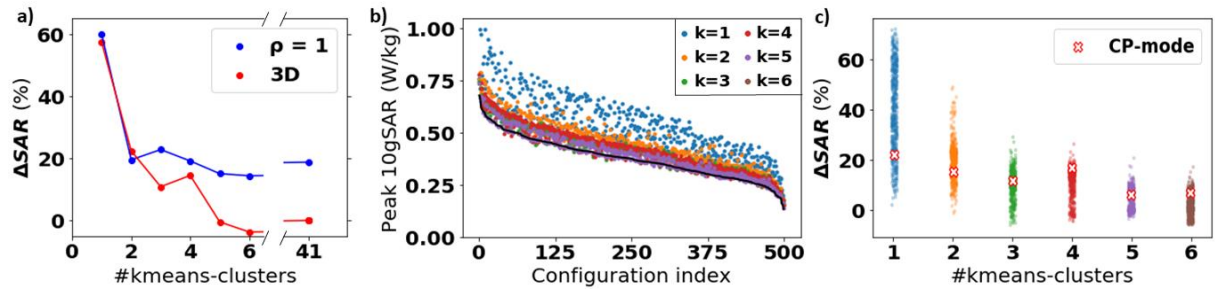

**Figure 3:** The reproducibility of the reference 10g-SAR-values for varying numbers of k-means clusters. **(a):** The error in worst-case 10g-SAR for Duke versus the number of clusters with all tissue densities set to $\rho = 1 \text{ g/cm}^3$ ("ρ = 1") and with densities included as a third parameter in the k-means clustering ("3D"). **(b):** The peak 10g-SAR values for 500 random pTx-configurations, versus the numbers of clusters in the Duke model using the density as a clustering parameter. The black line shows the full model-values for each shim, ordered by decreasing 10g-SAR. **(c):** The errors in the data in (b) for the different models for all shims, shown as percentages of the reference maximum 10g-SAR for each shim. The black line shows the average values and the corresponding standard-deviations for the different numbers of k-means clusters, and the crosses mark the errors when using CP-mode.

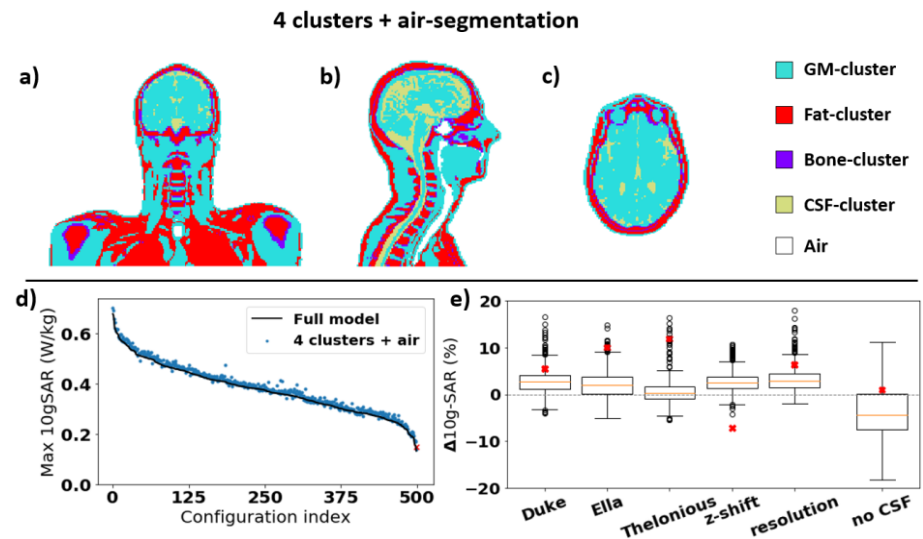

**Figure 4:** Simulation results for the '4 clusters plus air' model. **(a-c):** Three orthogonal slices of the simulated region of Duke, using the '4 clusters plus air' segmentation. The internal air is included in the same tissue-type as the background of the models (shown in white). **(d):** The simulated 10g-SAR values for 500 random $B_1$-shims for the original Duke model (black line) and for the '4 clusters plus air' segmentation of Duke (blue). The red cross indicates CP-mode. **(e):** Statistical representation of the relative errors in the peak-10g-SAR for all 500 shims. Results are shown for three different models (Duke, Ella, and Thelonious), for two variations of the simulation settings for the Duke-model: a 5 cm change in the position of Duke relative to the coil ("z-shift") and a 0.5 mm isotropic reduction of the maximum voxel size, and for an alternative segmentation (without the CSF-cluster), which can be segmented using the method presented by Milshteyn et al. (12). Red crosses indicate the values for CP-mode.

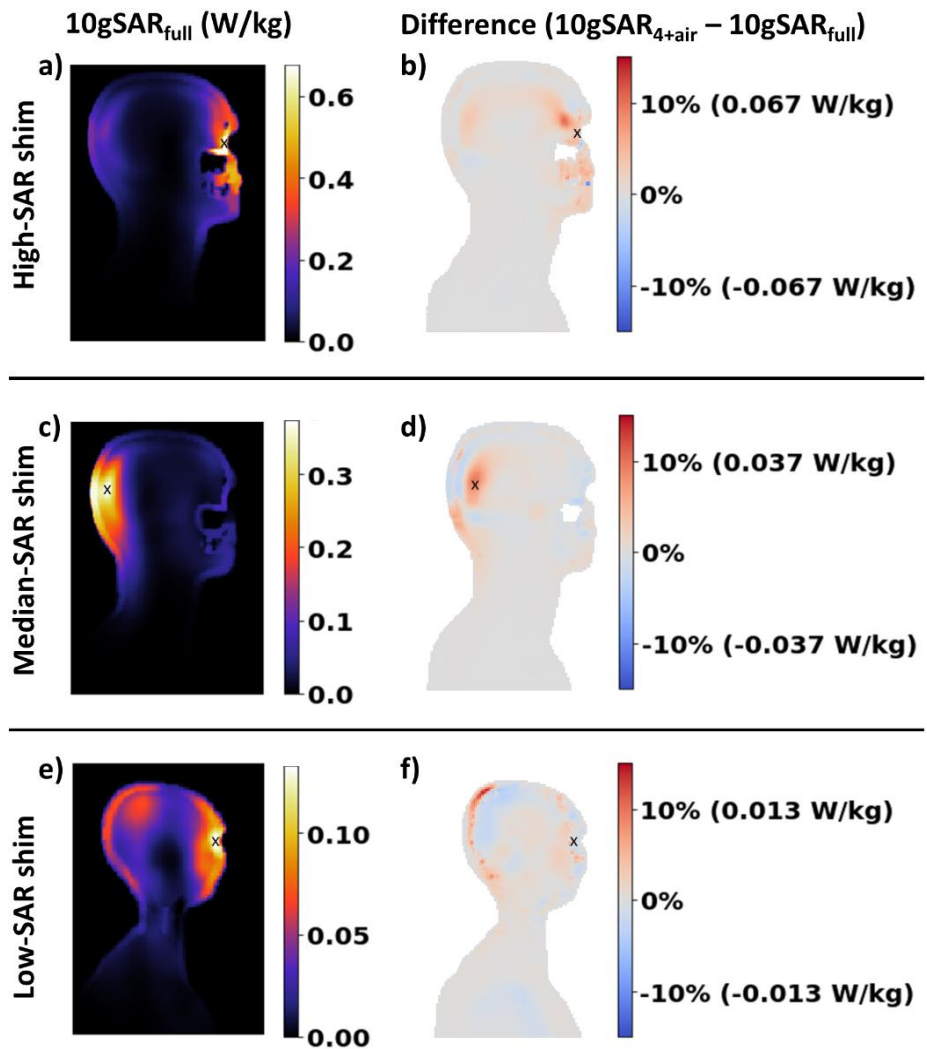

**Figure 5:** The in-plane difference in 10g-SAR between simulations using the '4 clusters plus air' model ("10g-SAR$_{4+air}$") and the full Duke model ("10g-SAR$_{full}$") for different $B_1$-shims. The presented slices are sagittal slices containing the voxel with the highest 10g-SAR-value in the full Duke model for the given shim. Small, black crosses ('x') indicate the locations with the highest 10g-SAR in the full model. Note that the magnitude of the errors in these single-slice distributions can be different than the values found in Figure 4d, which only compares the error in highest 10g-SAR-value between the two models instead of comparing the values on a voxel-by-voxel basis. **(a):** The in-plane simulated 10g-SAR distribution in the full Duke model when using the $B_1$-shim with the highest resulting 10g-SAR of the 500 random, normalized $B_1$-shims. **(b):** The errors in simulated 10g-SAR when using the '4 clusters plus air' model, relative to the peak-10g-SAR in the full model-simulation. Positive errors correspond to overestimations of the 10g-SAR. **(c-d):** The same as (a-b), when using the $B_1$-shim with the median 10g-SAR of the 500 shims. **(e-f):** The same for the $B_1$-shim with the lowest 10g-SAR of the 500 shims.

**Supporting Information**

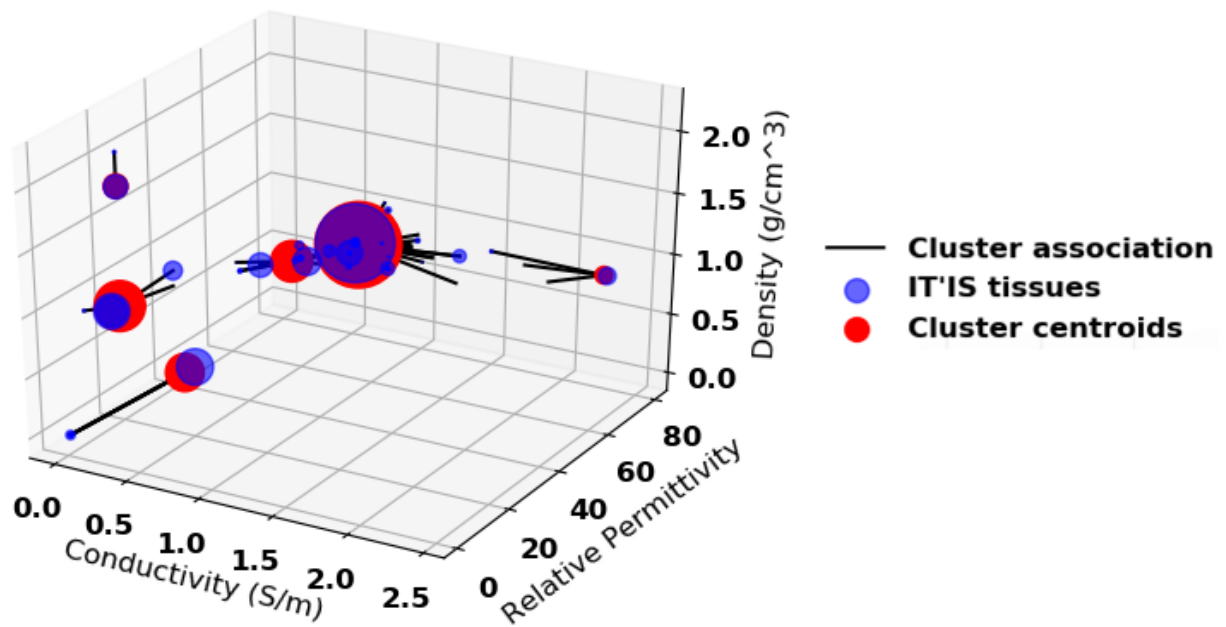

**Supporting Information Figure S1:** Example of three-parameter k-means clustering with *k=5* clusters. The dielectric properties of the tissues in the original Duke model in the IT'IS database are shown as blue spheres, the k-means centroids are shown as red spheres, and black lines indicate which centroid (cluster) each tissue is mapped to. The radius of each sphere is proportional to the volume of tissue represented by it.

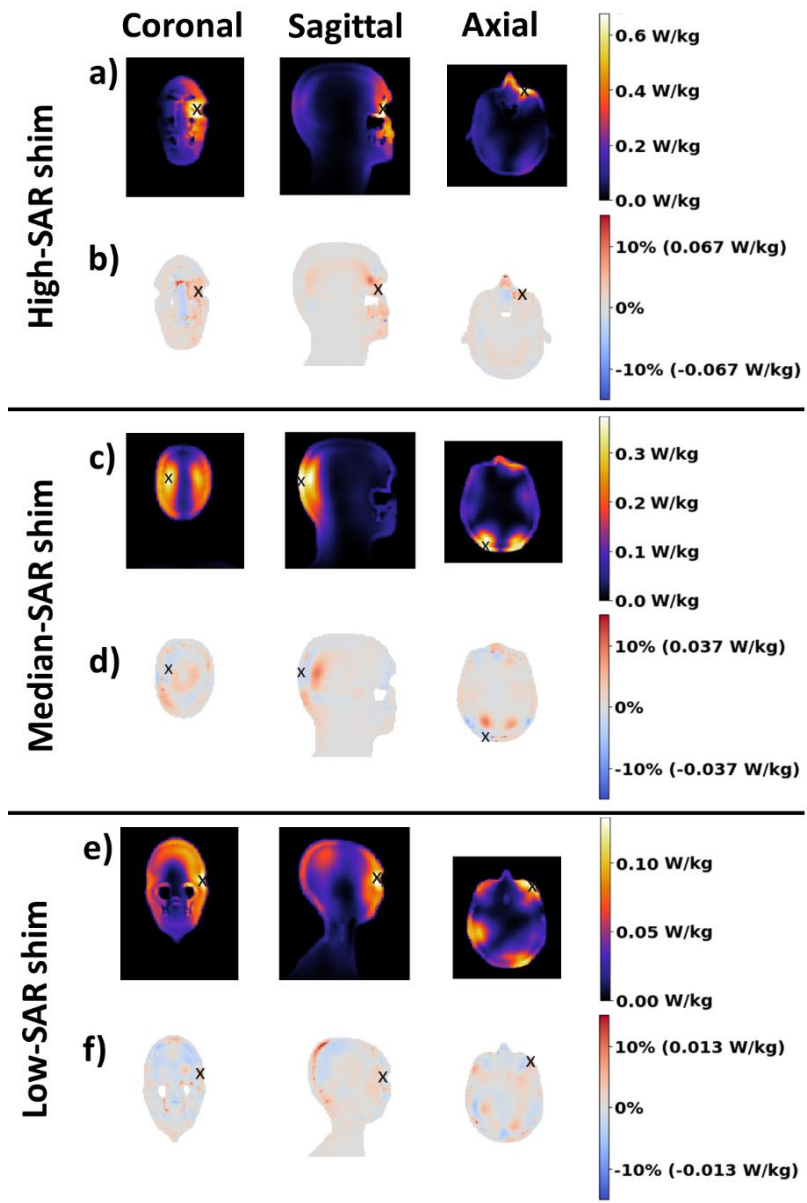

**Supplementary Information Figure S2:** The in-plane difference in 10g-SAR between simulations using the '4 clusters plus air' model ("$SAR_{4+air}$") and the full Duke model ("$SAR_{full}$") for different $B_1$-shims, displayed for three orthogonal slices (in addition to only the sagittal slices, shown in Figure 5 of the main text). The presented slices are the slices containing the voxel with the highest 10g-SAR-value in the full Duke model for each shim. This peak-SAR location is marked with an 'x'. **(a):** The in-plane simulated 10g-SAR distribution in the full Duke model when using the $B_1$-shim with the highest resulting 10g-SAR of the 500 random, normalized $B_1$-shims. **(b):** The errors in simulated SAR when using the '4 clusters plus air' model, relative to the peak-SAR in the full model-simulation. Positive errors correspond to overestimations of the SAR. **(c-d):** The same as (a-b), when using the $B_1$-shim with the median 10g-SAR of the 500 shims. **(e-f):** The same for the $B_1$-shim with the lowest SAR.

| Name | ρ (kg/m³) | σ (S/m) | ε (εᵣ) | V (L) | kMeans-5 | 4 + air |
|---|---|---|---|---|---|---|
| Esophagus Lumen | 1.16 | 0.00 | 1.00 | 2.0E-03 | 4 | Air |
| Bronchi lumen | 1.16 | 0.00 | 1.00 | 3.8E-03 | 4 | Air |
| Trachea Lumen | 1.16 | 0.00 | 1.00 | 2.6E-02 | 4 | Air |
| **Air 1** | **1.16** | **0.00** | **1.00** | **7.4E-02** | **4** | **Air** |
| Bone Marrow (Yellow) | 980.0 | 0.04 | 5.49 | 6.6E-03 | 2 | Fat |
| **Fat** | **911.0** | **0.12** | **11.29** | **8.5E-01** | **2** | **Fat** |
| SAT (Subcutaneous Fat) | 911.0 | 0.12 | 11.29 | 1.1E+00 | 2 | Fat |
| Tooth | 2180.0 | 0.16 | 12.36 | 8.6E-03 | 0 | Bone |
| **Bone (Cortical)** | **1908.0** | **0.16** | **12.36** | **5.1E-01** | **0** | **Bone** |
| Bone (Cancellous) | 1178.3 | 0.36 | 20.58 | 3.0E-01 | 2 | Fat |
| Lung | 394.0 | 0.47 | 21.83 | 1.2E+00 | 4 | Fat |
| Eye (Lens) | 1075.5 | 0.51 | 35.67 | 3.0E-04 | 1 | GM |
| Spinal Cord | 1075.0 | 0.60 | 32.25 | 1.5E-02 | 1 | GM |
| Nerve | 1075.0 | 0.60 | 32.25 | 9.2E-03 | 1 | GM |
| Commissura Anterior | 1041.0 | 0.62 | 38.58 | 4.0E-05 | 1 | GM |
| Commissura Posterior | 1041.0 | 0.62 | 38.58 | 8.0E-05 | 1 | GM |
| Brain (White Matter) | 1041.0 | 0.62 | 38.58 | 5.2E-01 | 1 | GM |
| Lymphnode | 1035.0 | 0.74 | 79.56 | 1.1E-03 | 1 | GM |
| Tendon\Ligament | 1142.0 | 0.76 | 45.63 | 6.3E-02 | 1 | GM |
| Trachea | 1080.0 | 0.80 | 41.78 | 2.4E-02 | 1 | GM |
| Bronchi | 1101.5 | 0.80 | 41.78 | 1.9E-03 | 1 | GM |
| Larynx | 1099.5 | 0.83 | 42.32 | 1.7E-02 | 1 | GM |
| Cartilage | 1099.5 | 0.83 | 42.32 | 3.2E-02 | 1 | GM |
| Salivary Gland | 1048.0 | 0.83 | 75.79 | 3.4E-02 | 1 | GM |
| Skin | 1109.0 | 0.90 | 40.94 | 6.9E-01 | 1 | GM |
| Tongue | 1090.4 | 0.98 | 55.02 | 7.2E-02 | 1 | GM |
| Mucous Membrane | 1102.0 | 0.98 | 54.81 | 5.1E-02 | 1 | GM |
| Muscle | 1090.4 | 0.98 | 54.81 | 5.6E+00 | 1 | GM |
| Thalamus | 1044.5 | 0.99 | 52.28 | 9.1E-03 | 1 | GM |
| Hypothalamus | 1044.5 | 0.99 | 52.28 | 6.0E-04 | 1 | GM |
| **Brain (Grey Matter)** | **1044.5** | **0.99** | **52.28** | **5.6E-01** | **1** | **GM** |
| Hippocampus | 1044.5 | 0.99 | 52.28 | 6.0E-03 | 1 | GM |
| Dura | 1174.0 | 0.99 | 44.20 | 1.2E-01 | 1 | GM |
| Pineal Body | 1053.0 | 1.08 | 59.47 | 1.0E-04 | 1 | GM |
| Hypophysis | 1053.0 | 1.08 | 59.47 | 9.0E-04 | 1 | GM |
| Thyroid Gland | 1050.0 | 1.08 | 59.47 | 4.9E-03 | 1 | GM |
| Intervertebral Disc | 1099.5 | 1.14 | 43.14 | 2.5E-02 | 1 | GM |
| Eye (Sclera) | 1032.0 | 1.21 | 55.02 | 4.1E-03 | 1 | GM |
| Esophagus | 1040.0 | 1.23 | 64.80 | 1.9E-02 | 1 | GM |
| Cerebellum | 1045.0 | 1.31 | 48.86 | 1.6E-01 | 1 | GM |
| Medulla Oblongata | 1045.5 | 1.31 | 48.86 | 4.1E-03 | 1 | GM |
| Midbrain | 1045.5 | 1.31 | 48.86 | 1.4E-02 | 1 | GM |
| Pons | 1045.5 | 1.31 | 48.86 | 1.5E-02 | 1 | GM |
| Eye (Cornea) | 1050.5 | 1.44 | 54.84 | 8.0E-04 | 1 | GM |
| Blood | 1049.8 | 1.58 | 61.06 | 1.5E-01 | 1 | GM |
| Eye (Vitreous Humor) | 1004.5 | 1.67 | 68.88 | 7.8E-03 | 3 | CSF |
| **Cerebrospinal Fluid** | **1007.0** | **2.46** | **68.44** | **2.6E-01** | **3** | **CSF** |

**Supporting Information Table S1:** An overview of the 47 different tissue types (with 41 different combinations of dielectric properties) in the original voxelized Duke model, sorted by the conductivities used in Sim4Life. The first four columns contain the original (adult) tissue properties as used in Sim4Life. From left to right, those are the name, the mass density, the conductivity, and the relative permittivity of the tissue types. The fifth column shows the total volume per tissue type in the voxelized model. The sixth and seventh column show the clustering of the tissues in the $k=5$ clusters k-means clustering ("kMeans-5"; see Table S2 for the dielectric properties of the 5 tissue clusters in that segmentation), where every number is the index of a cluster, and in the 4 clusters plus air-model ("4 + air"). The original tissues which all other tissues are mapped to in the 4 clusters plus air-model and their dielectric properties are highlighted in grey in the overview. Note that tissue types which do exist in the full Duke model but were not included in the bounding box in the simulations are not included in this overview. Also note that the lungs are mapped to different clusters in the $k=5$ cluster k-means segmentation and the '4 clusters plus air' segmentation due to the slightly different dielectric properties of the cluster centroids in those two models.

| Index | $\rho$ (kg/m³) | $\sigma$ (S/m) | $\epsilon$ ($\epsilon_r$) | Allocation in "4 + air"-model |
|---|---|---|---|---|
| 0 | 1912 | 0.156 | 12.4 | Bone (Cortical) |
| 1 | 1085 | 0.965 | 52.1 | Brain (Grey Matter) |
| 2 | 947 | 0.149 | 12.5 | Fat |
| 3 | 1007 | 2.432 | 68.5 | Cerebrospinal Fluid |
| 4 | 362 | 0.435 | 20.1 | Air 1 |

**Supporting Information Table S2:** The dielectric properties of the $k=5$ clusters in the k-means segmentation. Indices refer to the indices in the 6th column ("kMeans-5") in Supporting Information Table S1. The final column indicates the tissue types in the IT'IS database from which the dielectric properties were used in the final '4 clusters plus air'-model.

**a) Duke**

|  | With CSF | Without CSF |
|---|---|---|
| Error: worst-case 10g-SAR | 1.0 % | −8.8 % |
| Displacement: worst-case 10g-SAR | 0.0 cm | 20.7 cm |
| Errors in 500 shims | 2.8 ± 2.8 % | −4.1 ± 5.9 % |
| 99% certainty margin | 12.3 % | 15.7 % |

**b) Ella**

|  | With CSF | Without CSF |
|---|---|---|
| Error: worst-case 10g-SAR | 1.6 % | −6.4 % |
| Displacement: worst-case 10g-SAR | 0.0 cm | 6.0 cm |
| Errors in 500 shims | 2.3 ± 3.2 % | −3.8 ± 7.2 % |
| 99% certainty margin | 11.4 % | 19.3 % |

**Supporting Information Table S3:** Comparison if the accuracy of SAR simulations when using the '4 clusters plus air'-model with CSF and when excluding the CSF from the model.